# Remote Access Mössbauer Spectrometry

**A. A. Silaev (Jr.), S. K. Godovikov, E. B. Postnikov, V. V. Radchenko, and A. A. Silaev (Sr.)**
*Skobel'tsyn Institute of Nuclear Physics, Moscow State University, Moscow, 119991 Russia*
*e-mail: postn@dec1.sinp.msu.ru*

**Abstract**—The authors' design for a Mössbauer spectroscopy unit accessed via the Internet is presented. The spectrometer's configuration, operational peculiarities, key specifications, and remote access procedures are described. More detailed educational, scientific, and technical information can be found at the project's official website http://efmsb.sinp.msu.ru.

## INTRODUCTION

Mössbauer spectroscopy is employed in many different fields of human endeavor, from validating the general theory of relativity to studying the surface of Mars [1, 2]. It is an integral part of practical training in physics (nuclear physics, solid state physics, spectroscopy, etc.), since it allows us to investigate unique properties of a microcosm, e.g., the ultrafine structure of the energy levels of an atomic nucleus [1], and to become proficient in modern methods for studying the condensed state of matter [3–6].

Unfortunately, many scientific and industrial laboratories and institutes not only cannot afford to buy modern devices and equipment, they cannot maintain the instrumental base they already have in working condition. At the same time, most of the laboratory setups at Moscow State University's Institute of Nuclear Physics (INP MSU) have been updated and computerized over the last 15 years, thanks to the unique developments of the institute's specialists. The Mössbauer spectrometry laboratory in particular has been renovated substantially [7].

In 2011, a system for remote access to our unique in-house designed Mössbauer spectrometer was developed at INP MSU under a state contract with the RF Ministry of Education and Science as part of the federal target program Developing the Infrastructure of Nanoindustry in the Russian Federation. The aim of our work was to provide an opportunity to perform scientific research in real time with an actual source for any user anywhere on RF territory. In addition, it was possible for novice users to preliminarily study the foundations of Mössbauer spectroscopy.

The concept and architecture of the Mössbauer complex at INP MSU were developed in light of worldwide experience in the field of interactive remote studies according to the latest standards for online educational resources. The complex's scientific research value can be attributed to the possibility of gaining remote access to nondestructive Mössbauer testing of the parameters of crystalline, polycrystalline, and amorphous objects, including nanostructures, ensembles of nanoparticles, powders, and single crystals. Remote access is granted to the control unit of the spectrometer's vibrator (frequency and amplitude monitoring), and to the system for the automated remote replacement of specimens. The parameters of the amplifier and analyzer can be also set.

Despite the great number of advantages of remote access to a Mössbauer spectrometer for educational and research purposes, this problem had never been solved anywhere in the world. For it to be implemented, a number of engineering and technological difficulties were successfully overcome at INP MSU. These included reequipping the spectrometric devices with a digital instead of analog control system and improving its immunity to external factors (temperature); combining different modules (signal analyzers, vibrator generator, high-voltage power source) into one block with a common control system; mating the experimental facility with a PC and ensuring complete control over it via a computer interface; developing a digital functional generator to produce an actuating signal for the vibrator that is capable of specifying the frequency and amplitude of the signal from the computer, etc.

## CONCEPT OF REMOTE CONTROL

Specialists all over the world today have gained vast experience in implementing remote training systems. This concept has become increasingly relevant in the last few years, since it allows students to acquire high-quality education without substantial cost in time and materials. For example, nearly 1 million people underwent remote training in more than half of the United States' universities in 2004 [8].

Internet systems based on using unique devices and equipment that operate in real time allow us not only to expand the circle and geography of trainees, but also to enable research teams to conduct remote experi-mental studies. Such systems are especially important in training specialists for key high-tech industries, since full-scale facilities require enormous capital outlays and are, as a rule, unique. Such opportunities are offered by the interactive educational and scientific complex based on the unique Mössbauer spectrometer at INP MSU, which operates in the remote access mode.

A remote access laboratory is any educational, scientific, or industrial laboratory whose equipment can be controlled remotely via a computer connected to the Internet. The key aims of remote access laboratories are to be part of the training courses in traditional student practicals; to help expand the potentialities of the educational process; and to increase the quality of remote training in the engineering and technical sphere, which requires the practical skills of working in a laboratory. The Mössbauer educational and scientific complex at INP MSU is also intended for independent distant spectroscopic studies performed by the specialists working in different fields of science and engineering.

The key functional capabilities of remote access laboratories are guided by this objective as well, as is reflected in the concepts they bring to life. Among the functional capabilities of such laboratories are the use of visual methods and multimedia, interactive methods, high-quality graphic computer visualization, and the high scientific potential and competitive specifications of their laboratory equipment.

These capabilities should not only fully compensate for the drawbacks related to a student or researcher not being in a real laboratory; they should also introduce modern trends into the educational process that improve its efficiency but are unfortunately not affordable in traditionally organized practical work. These include the possibility of a large group of students staging one experiment simultaneously, or performing an experiment using a mathematical model proposed by the experimental facility, or conducting an experiment that requires the long-term accumulation of statistical data under conditions that are more comfortable for the researcher, i.e., without his or her around-the-clock presence in a laboratory (or any presence at all).

The simplest concept for a remote access laboratory is a functional auxiliary module in a conventional laboratory setup that connects the experimental equipment with an Internet server and thus allows the tracking and monitoring of actual experiments from a remote location.

A more complex and widely used concept for a remote access laboratory that includes the completely independent staging of experiments using laboratory equipment through an Internet-connected computer requires not only the experimental equipment to be mated to the computer server but also the full automation of the experimental laboratory setup. The laboratory equipment can then be used more efficiently

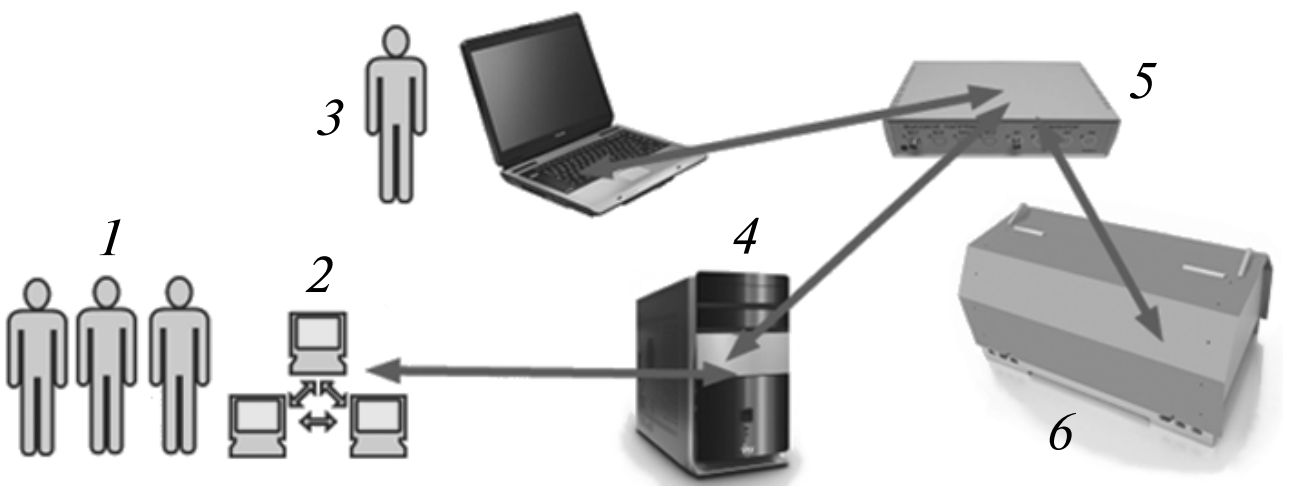

**Fig. 1.** Schematic of remote access to the spectrometer: (*1*) remote users; (*2*) Internet; (*3*) local user; (*4*) spectrometer server, (*5*) electronic unit; (*6*) Mössbauer spectrometer.

(daily, around the clock), and the permanent presence of laboratory personnel is not required.

The remote access Mössbauer spectrometer at INP MSU was created to develop the concept of staging fully independent remote experiments without the participation of laboratory personnel, using remote equipment via an Internet-connected computer (Fig. 1). Its mechanical elements are fully automated, and there are corresponding software and information modules for studying the operation of the entire system.

## DESCRIPTION OF THE SETUP

The complex consists of hardware elements that include the mechanical components of the spectrometer, its electronic unit, and computer equipment. The software elements of the complex include a remote access server, a remote access client, a Mössbauer spectrometer simulator, and a website (as the main communications resource) from which the user can start his or her work with the complex.

The mechanical components of the spectrometer were developed to provide remote access to users in daily and around-the-clock modes without the participation of a spectrometer operating crew. For such functionality to be ensured, none of the spectrometer elements are monitored mechanically, and they are controlled entirely from the computer. One of the key elements of the spectrometer that allows it to operate in the remote access mode is a robotized system for replacing samples (Fig. 2). When the complex is operated, up to 25 samples (for both educational and research work) are loaded into the system, and the user can work remotely with his or her specimens at any time.

The mechanical units of the complex are intended for precision studies of specific material samples with the help of Mössbauer spectroscopy. The accuracy of the mechanical parts' operation is ensured through the application of modern methods for the digital generation of actuating signals and monitoring the error signal from the measuring coil of the vibrator.

The spectrometer electron unit consists of the following modules (Fig. 3):

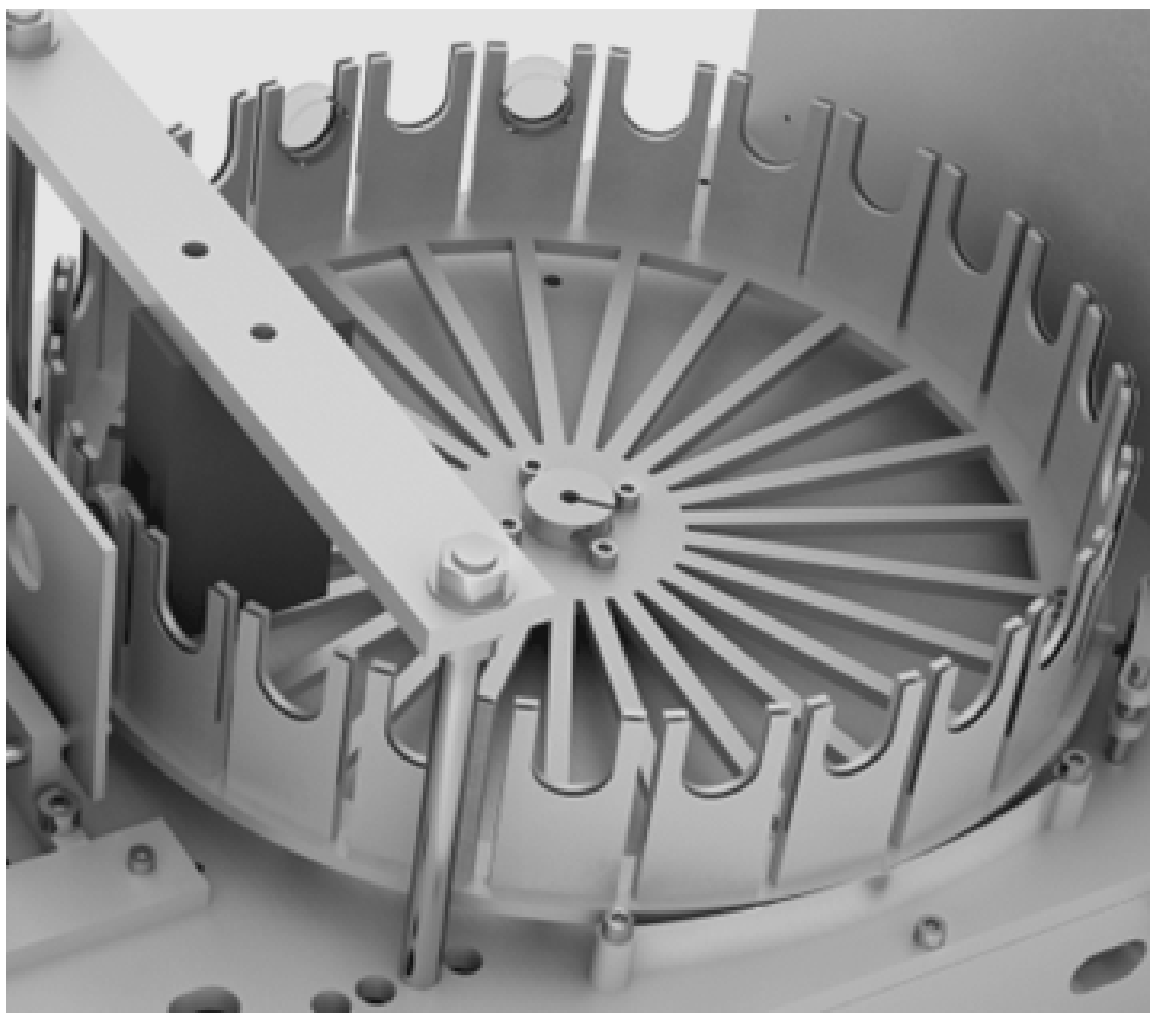

**Fig. 2.** Rotary system for replacing sample.

—low-voltage power supply (+12 V, +5 V, −5V, +3.3 V);

—high-voltage power supply for the detector (a gas-filled proportional counter) 0–2100 V;

—a vibrator controller that provides Doppler modulation with a constant rate of acceleration of up to 40 mm $s^{-1}$ in the 1–50 Hz range of frequencies;

—an analyzer of detector signals with a 10-bit ADC that is responsible for multiplying signals in the 1–8 range.

Communication with the computer is established via a USB-2 port at a maximum counting rate of 200 kHz.

From the standpoint of its overall dimensions, the spectrometer is more compact than its previous non-automated version, and can be conveniently installed in any laboratory. The mechanical and electronic block of the setup measures 700 × 400 × 320 mm and weighs 55 kg (including 4 mm of lead protection from the radiation of the radioactive source).

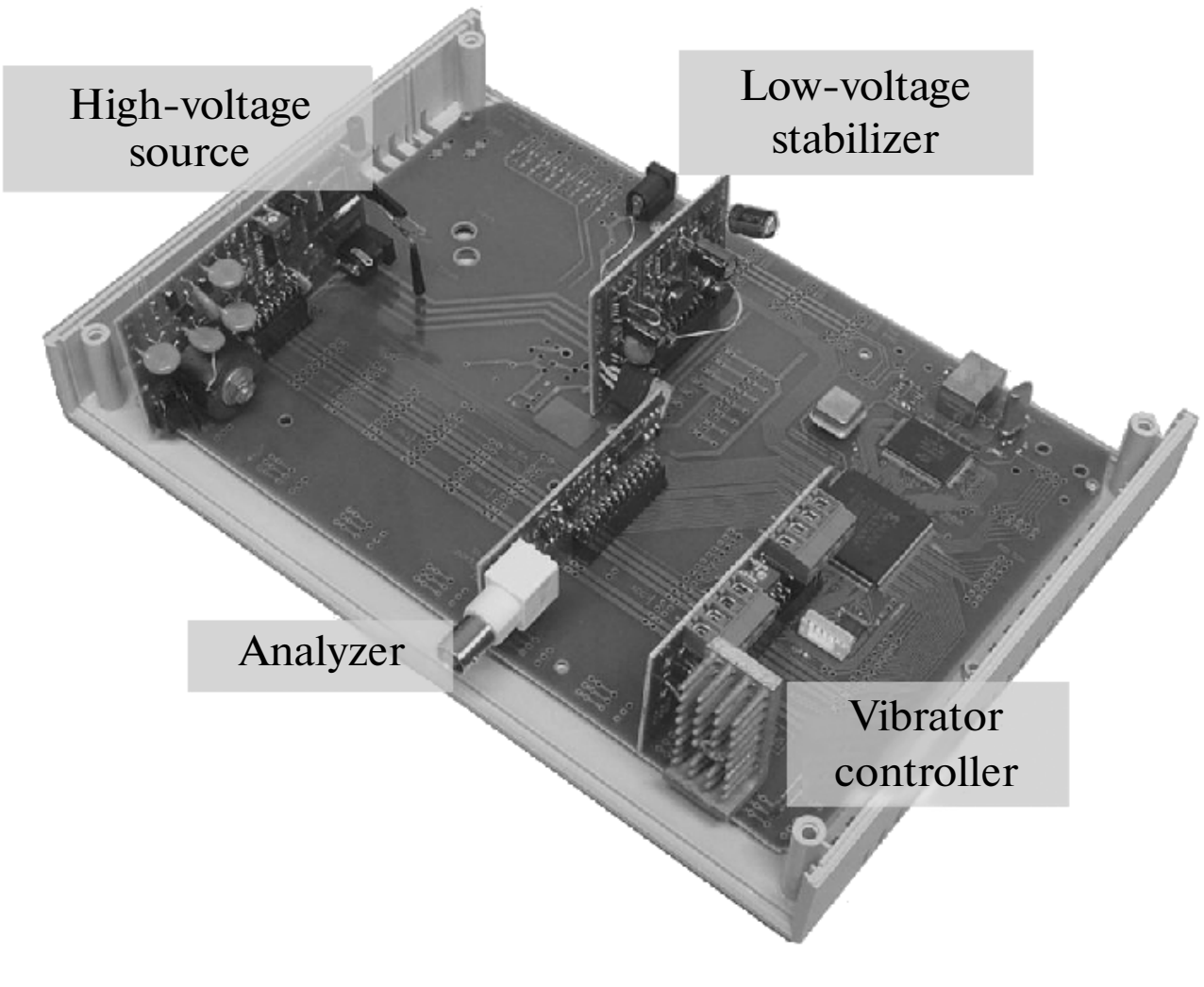

**Fig. 3.** Electronic unit (26 × 16 cm).

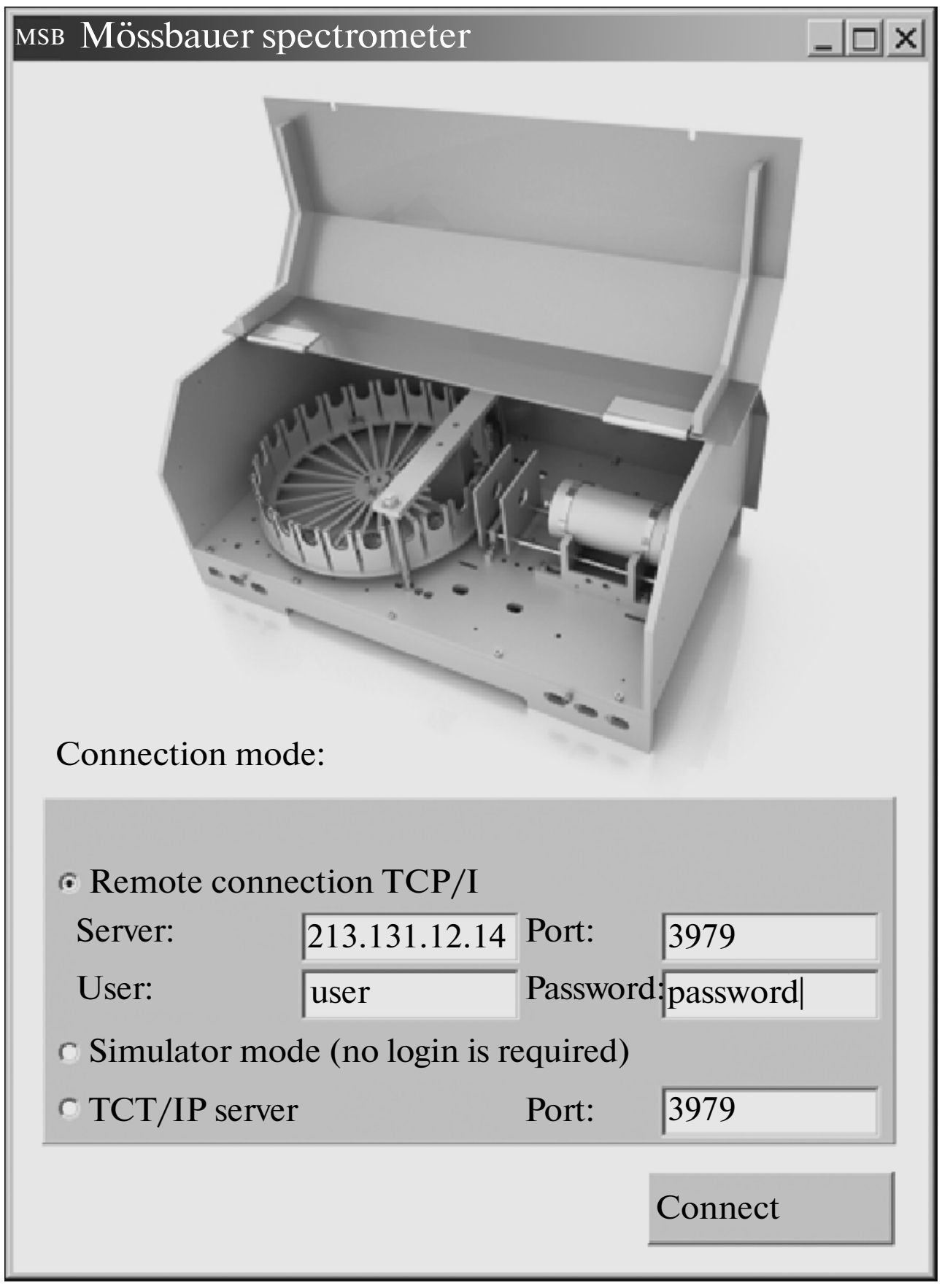

**Fig. 4.** Selection of the mode of connection to spectrometer.

The server program (characterized by a wider set of control potentialities) and the client program contained in the software of the complex allows users to obtain remote access for staging remote experiments (Figs. 4 and 5). The simulator of the Mössbauer spectrometer allows users to acquaint themselves with the spectrometer's configuration and operation, and to conduct virtual experiments (Fig. 6). The website contains interactive electronic training modules, a simulator, all of the required software, and a system of user registration for remote operation.

## REMOTE ACCESS TO THE SPECTROMETER

Since the remote access Mössbauer spectrometer is fully automated and can be operated from any computer connected to the Internet, the key problem in organizing the educational and research program is distributing the time of access to the equipment. For training in the remote access mode to be organized, a group of users headed by a supervisor must register. To get permission to work with remote equipment, the supervisor must file an application at the Mössbauer spectrometer website. Once the corresponding procedures are complete, he or she receives logins and passwords for all users of the group and a timetable or time quotas for free access to the equipment during its downtime.

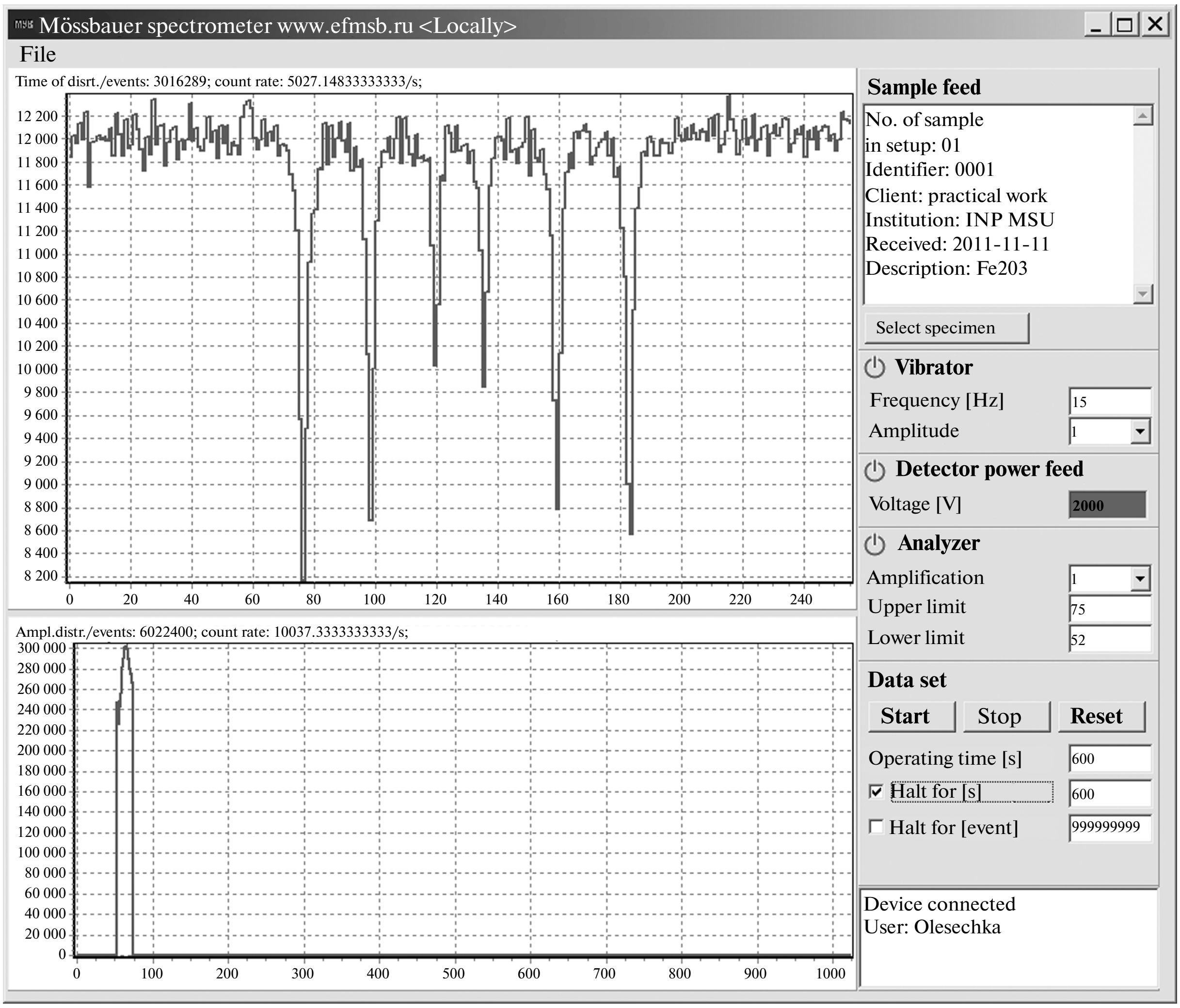

**Fig. 5.** Mössbauer spectrum (top) and amplitude spectrum (bottom) obtained from remote operation with the spectrometer.

To work with the remote equipment, the users must acquaint themselves with all of the site's electronic tutorials, answer all of its test questions, and pass a virtual test for working with the Mössbauer spectrometer's simulator. If there is no possibility of getting remote access to the equipment, the simulator can substitute for the staging of an actual experiment since it fully simulates the actual equipment and generates data using results obtained on it (Fig. 6).

The efforts of a registered user are aimed at performing experiments on remote equipment. The user gains access to the equipment at the time specified upon registering and conducts a remote experiment using the remote access client program.

Research work on the remote access spectrometer requires additional coordination between the researcher and the attending crew of the spectrometer. This is due to the need for delivering research samples from the user to the spectrometer personnel. The problem of delivery is solved through correspondence between the researcher and the attending crew. The samples can be delivered personally or by mail. The researcher's work with the complex is no different from the work of a student since one and the same software is used.

## CONCLUSIONS

A fully automated Mössbauer spectrometer controlled from a personal computer has been developed at INP MSU. It is characterized by compactness and a highly precise accuracy of measurement. The spectrometer can be remotely accessed through an Internet-connected computer. This scientific and educational setup has no analogs anywhere in the world.

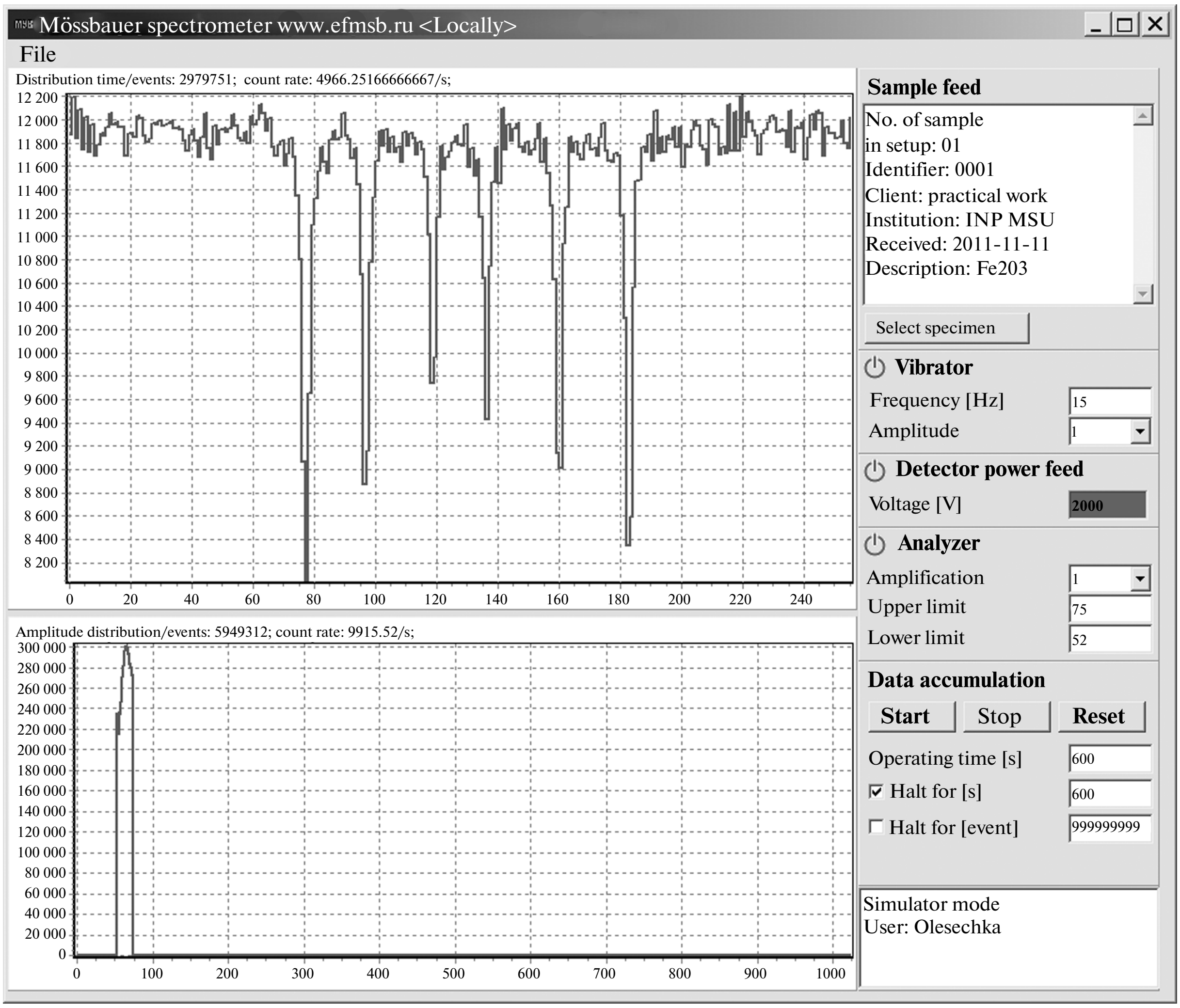

**Fig. 6.** Results from remote operation using our mathematical model of the Mössbauer spectrometer (simulator): Mössbauer spectrum (top); amplitude spectrum (bottom).

## ACKNOWLEDGMENTS

The study was supported by the RF Ministry of Education and Science, state contract no. 16.647.12.2044 of July 15, 2011. The results from implementing our project can be found in the national nanotechnology network.

## REFERENCES

1 1. Wertheim, G.K., *Mössbauer Effect: Principles and Applications*, New York: Academic, 1964.

2. Rodionov, D.S., Mössbauer Spectrometer for Analyzing Ferrum Mineralogy on the Mars Surface, *Cand. Sci. (Physmath.) Dissertation*, Moscow: Space Res. Inst., 2006.

3. Ovchinnikov, V.V., *Messbauerovskie metody analiza atomnoi i magnitnoi struktury splavov* (Mossbauer Analytical Methods for Atomic and Magnetic Structure of Alloys), Moscow: Fizmatlit, 2002.

4. Vilkov, L.V. and Pentin, Yu.A., *Fizicheskie metody issledovaniya v khimii. Rezonansnye i elektroopticheskie metody* (Physical Research Methods in Chemistry. Resonance and Electrooptical Methods), Moscow: Vysshaya shkola, 1989.

5. Kamnev, A.A., *J. Molec. Struct.,* 2005, vols. 744–747, pp. 161–167.

6. *Industrial Applications of the Mossbauer Effect*, Long, G.J. and Stevens, J.G., Eds., New York: Plenum Press, 1986.

7. Popov, Yu.V., Radchenko, V.V., and Persiantsev, I.G., http://www.phys.msu.ru/rus/about/sovphys/ISSUES-2006/1%2848%29-2006/praktikum/

8. *National Tech. Univ. Bull.*, National Technological Univ, Minneapolis, 2004, vol. 16, no. 1. http://www.ntu.edu/Ac/Bulletin/documents/00Bulletin 2004-05No1July.pdf

*Translated by L. Borodina*

SPELL: 1. ok